\documentclass[12pt]{article}
\usepackage{array,xspace,multirow,hhline,tikz,colortbl,tabularx,booktabs,fixltx2e,amsmath,amssymb,amsfonts,amsthm}
\usetikzlibrary{arrows}
\usetikzlibrary{shapes}
\usepackage{algorithm,algorithmic}
\usepackage{eqparbox}
\usepackage{verbatim,ifthen}
\usepackage{pifont} 
\usepackage{setspace}
\usepackage{graphicx}
\usepackage{caption} 
\usepackage{pgfplots}
\usepackage{bbm}
\usepackage{wrapfig} 
\usepackage{natbib} 
\usepackage{ifthen}
\usepackage{calrsfs,mathrsfs}
\usepackage{bbding,pifont} 
\usepackage{pgflibraryshapes} 
\usepackage{url}  
\RequirePackage[bookmarks, bookmarksopen=true, plainpages=false, pdfpagelabels, pdfpagelayout=SinglePage, breaklinks = true]{hyperref}

\usepackage{geometry}
\usepackage{abstract}
\usepackage{tcolorbox}

 \usepackage[bottom,flushmargin,hang,multiple]{footmisc}

\usepackage{gensymb} 
 
\usepackage{array,xspace,hhline,tikz,colortbl,tabularx,fixltx2e,amsmath,amssymb,amsfonts,amsthm}
\usepackage{verbatim,ifthen}
\usepackage{pifont}
\usepackage{ifthen}
\usepackage{pgflibraryshapes}
\usepackage[multiple]{footmisc}
\usepackage{nicefrac}
\usetikzlibrary{arrows}
\usepackage{ltxtable}
\usepackage{longtable}

	\usepackage{varioref}

\tikzset{
  jumpdot/.style={mark=*,solid},
  excl/.append style={jumpdot,fill=white},
  incl/.append style={jumpdot,fill=black},
  rexcl/.append style={jumpdot,color=red,fill=white},
  rincl/.append style={jumpdot,fill=black,color=red},
}

	\usepackage{varioref} 
		\usepackage{subfigure}

\definecolor{light-gray}{gray}{0.9}

\newtheorem{innercustomgeneric}{\customgenericname}
\providecommand{\customgenericname}{}
\newcommand{\newcustomtheorem}[2]{%
  \newenvironment{#1}[1]
  {%
   \renewcommand\customgenericname{#2}%
   \renewcommand\theinnercustomgeneric{##1}%
   \innercustomgeneric
  }
  {\endinnercustomgeneric}
}

\newcustomtheorem{customthm}{Theorem}
\newcustomtheorem{customlemma}{Lemma}
\newcustomtheorem{customproposition}{Proposition}

	\newtheorem{example}{Example}
	
\usepackage{enumerate}

\hypersetup{
	colorlinks=true,
	linkcolor=blue,
	filecolor=blue,      
	urlcolor=blue,
	citecolor=blue
}

\usepackage[normalem]{ulem}
	\newcommand\eat[1]{}

	\usepackage{enumitem}
	\setenumerate[1]{label=\rm(\it{\roman{*}}\rm),ref=({\it\roman{*}}),leftmargin=*}
	\newlength{\wordlength}

	\newcommand{\eqclass}[2][]{\ifthenelse{\equal{#1}{}}{[#2]}{[#2]_{\sim_{#1}}}}

\usepackage{enumitem}
\setenumerate[1]{label=\rm(\it{\roman{*}}\rm),ref=({\it\roman{*}}),leftmargin=*}

\newcommand{\nbh}[1][]{
	\ifthenelse{\equal{#1}{}}{\nu}{\nu(#1)}
}

\newcommand{\cstr}[1][]{
	\ifthenelse{\equal{#1}{}}{\mathscr S}{\cstr(#1)}
}

\newcommand{\choice}[1][]{
	\ifthenelse{\equal{#1}{}}{\mathit{C}}{\choice(#1)}

		\newcommand{\ml}[1][]{\ensuremath{\ifthenelse{\equal{#1}{}}{\mathit{ML}}{\mathit{ML}(#1)}}\xspace}
		\newcommand{\sml}[1][]{\ensuremath{\ifthenelse{\equal{#1}{}}{\mathit{SML}}{\mathit{SML}(#1)}}\xspace}
		\newcommand{\sd}[1][]{\ensuremath{\ifthenelse{\equal{#1}{}}{\mathit{SD}}{\mathit{SD}(#1)}}\xspace}
		\newcommand{\rsd}[1][]{\ensuremath{\ifthenelse{\equal{#1}{}}{\mathit{RSD}}{\mathit{RSD}(#1)}}\xspace}
		\newcommand{\rd}[1][]{\ensuremath{\ifthenelse{\equal{#1}{}}{\mathit{RD}}{\mathit{RD}(#1)}}\xspace}
		\newcommand{\st}[1][]{\ensuremath{\ifthenelse{\equal{#1}{}}{\mathit{ST}}{\mathit{ST}(#1)}}\xspace}
		\newcommand{\bd}[1][]{\ensuremath{\ifthenelse{\equal{#1}{}}{\mathit{BD}}{\mathit{BD}(#1)}}\xspace}
		\newcommand{\pc}[1][]{\ensuremath{\ifthenelse{\equal{#1}{}}{\mathit{PC}}{\mathit{PC}(#1)}}\xspace}
		\newcommand{\dl}[1][]{\ensuremath{\ifthenelse{\equal{#1}{}}{\mathit{DL}}{\mathit{DL}(#1)}}\xspace}
		\newcommand{\ul}[1][]{\ensuremath{\ifthenelse{\equal{#1}{}}{\mathit{UL}}{\mathit{UL}(#1)}}\xspace}

			\newcommand{\indiff}{\ensuremath{\sim}}}

			\usepackage{palatino}

\begin{document}

%
%
%
%
%
%
%
%


	
	\title{How to make the toss fair in cricket?} 
	



	
	\author{Haris Aziz\thanks{UNSW Sydney, Australia. Email: \href{mailto: haris.aziz@unsw.edu.au}{haris.aziz@unsw.edu.au}}  \\\small{\href{http://www.cse.unsw.edu.au/~haziz/CricketToss.pdf}{Latest version here}} \vspace{2mm} \\}

\date{27th Nov, 2021} 


	\maketitle

		\begin{abstract} 
			
			In the sport of cricket, the side that wins the toss and has the first choice to bat or bowl can have an unfair or a critical advantage. The issue has been discussed by International Cricket Council committees, as well as several cricket experts. In this article, I outline a method to make the toss fair in cricket. The method is based on ideas from the academic fields of game theory and fair division. 
	\end{abstract}

		
		\section{Introduction}
		
Do you remember when your heart sank when your team lost the toss of a crucial match on a tricky pitch? Or do you recall the annoying moment when the rival fans downplayed your team’s victory by pointing out the toss outcome? 

There is no doubt that the outcome of a cricket match can be affected by the toss of a coin~\citep{Mong18a}. On a fresh lush pitch, the test team being sent out to bat first may already be at a disadvantage as the juicy pitch provides ideal conditions for seam and swing. On a dry dusty pitch, the team that bats last on the fifth day faces a significant disadvantage as the pitch will have greatly deteriorated over the course of the match. 

The winner of the toss can also get an undue advantage in T20 matches especially in dewy conditions in the evenings. The trend has been noticed in various T20 tournaments including PSL~\citep{Raso21a}. During the ICC T20 World Cup 2021, 12 out of 13 matches in Dubai were won by the chasing team. Simon Burnton from The Guardian wrote that ``\emph{every evening match has seen 22 highly skilled sportspeople spend several hours straining to see if they can have a greater impact on the result than the momentary flight of a small metal disc before the action began, and in general they have failed}''~\citep{Burn21a}. When huge stakes are at play (such as the final of an ICC World Cup or the ICC Test Championship), it seems highly unreasonable for the outcome of the match to be affected by the outcome of the toss. When two evenly matched teams play, you want an exciting and evenly matched game.  \citet{Chop13a} pointedly asked: 

\begin{quote}
\emph{``Would football tolerate a toss of a coin making one side play with only 10 players? In swimming or Formula One could chance decide pole position? Would every boxing match begin with the tossing of a coin which would permit one of the two boxers a chance to land a couple of hard rights as the bout began? I don't think so.''}
\end{quote}

\section{The Toss, Propose and Choose Method}

The issue of the unreasonable impact of the toss has been discussed for years~\citep{SoWi16a}. It is important enough that the ICC has considered searching for fairer solutions~\citep{Cric18a}. So how can we ensure that the toss continues to be meaningful but doesn’t have an unfair effect on the outcome of the match? 
I have a  proposal to keep the toss in cricket but make it fair. I propose a minimal adaptation to the rules of cricket. I call the proposal the Toss, Propose and Choose method.

\begin{tcolorbox}
	\textbf{The Toss, Propose and Choose method. }	
	\begin{enumerate}
		\item[] \textbf{TOSS}: The toss takes place as per tradition. [Let us call the captain who wins the toss the ‘Lucky Captain’ and the captain who loses the toss the ‘Unlucky Captain’]. 
		\item[] \textbf{PROPOSE}: The Unlucky Captain judges whether bowling or batting first is disadvantageous and estimates the marginal impact of bowling versus batting first in terms of runs. Say the estimated number of runs is $b$. The Unlucky Captain proposes the following two options to the Lucky Captain:
\begin{enumerate}
	\item[]Option 1. Take the advantageous turn but give $b$ bonus runs to the other team
\item[] Option 2. Take the disadvantageous turn but get $b$ bonus runs. 
\end{enumerate}
	
\item[] \textbf{CHOOSE}: The Lucky Captain (who won the toss) then chooses either option 1 or 2.
	\end{enumerate}
	
	\end{tcolorbox}

\begin{figure}[h!]
	
	\centering
	\includegraphics[scale=0.03]{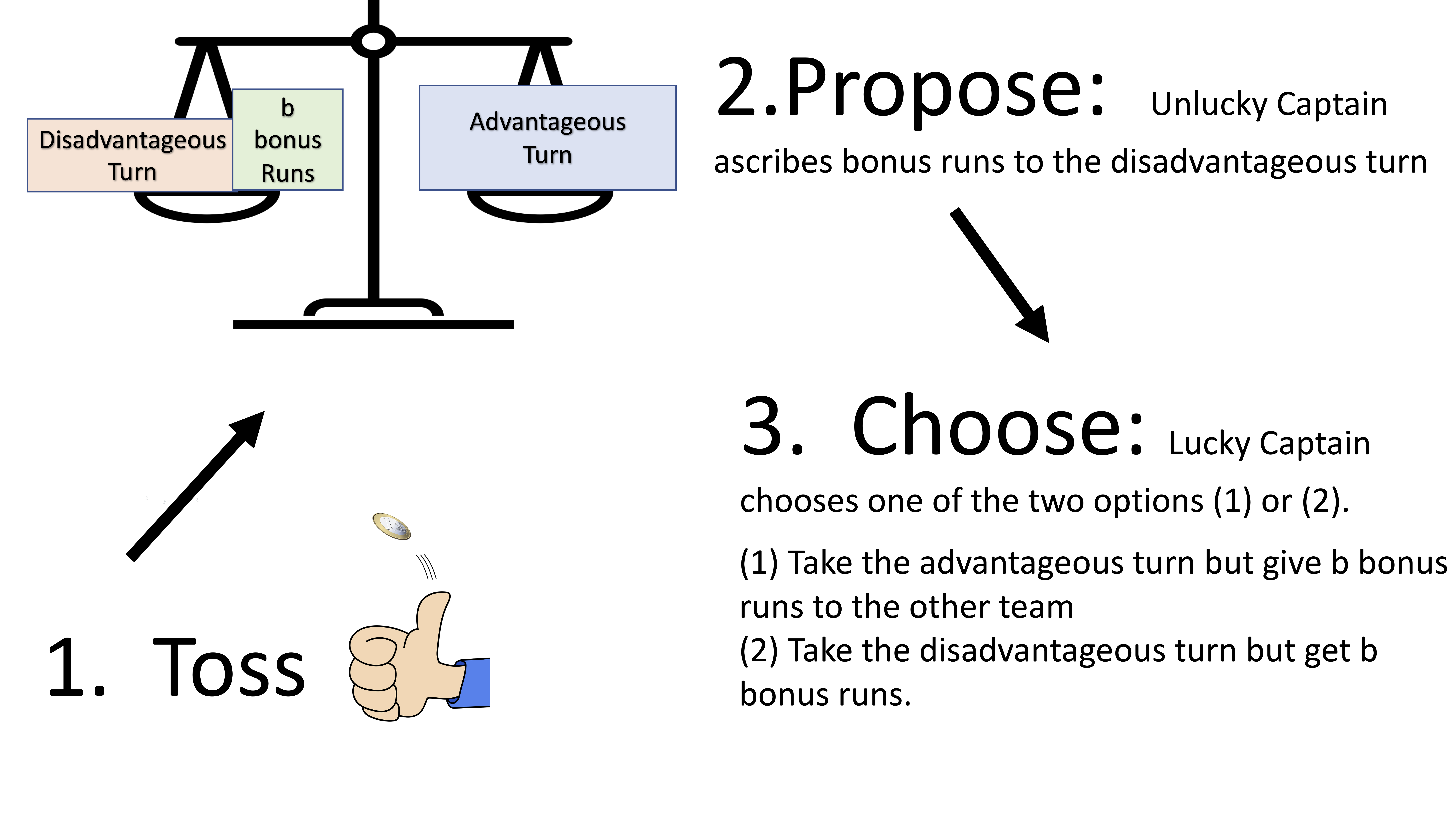}
	\caption{A graphical illustration of the Toss, Propose and Choose method. }
	\label{fig:toss}
	\end{figure}

Figure~\ref{fig:toss} gives a graphical illustration of the Toss, Propose and Choose method

\begin{example}
Let us illustrate the proposed method with the help of an example. Suppose Australia is playing New Zealand in the ICC Test Championship Final. We do not wish for a championship spanning multiple years to be eventually heavily influenced by the toss outcome. The coin is tossed and Australia’s captain wins. At this point, the New Zealand captain is asked to give two options to the Australian captain. New Zealand factors in the playing conditions and team lineups, and feels that they will gain a net 50 run advantage if they bowl first. In view of this, the Australian captain gets two options: bowl or bat first but with the team batting first getting 50 additional runs in the extras column. 
\end{example}

Now let us logically explain why the proposed method is fair to both parties.  The Unlucky Captain (who loses the toss) gets a full opportunity to ensure that the possible options (1 and 2) weigh up equally, so irrespective of which option is chosen, s/he does not feel unlucky. More precisely, the Unlucky Captain has no envy and does not wish to switch between the two options. The Lucky Captain gets the final say in choosing the preferred option among 1 and 2. Hence, the Lucky Captain also feels that their team is not disadvantaged. Voila! The method I am advocating has solid foundations as it can be viewed as a suitable adaptation of the classical Divide and Choose method for fairly dividing a divisible resource~\citep{BrTa96a}. Just as Divide and Choose, my method does leave scope for the New Zealand Captain to be strategic. For example, if Australia’s captain habitually prefers bowling first, the New Zealand  captain can increase the bonus runs from 50  to 70. However, even under strategic behaviour, the outcome remains fair (envy-free). 

Let me give another perspective of why the proposed method is fair. Suppose bowling first versus batting are compared on a balanced weight scale by the unlucky captain. The unlucky captain is concerned that the lucky captain is going to select the heavier side. The unlucky captain has the chance to put sufficient weight / bonus runs on the lighter side to ensure both sides have the same weight. In that case the Unlucky Captain does not need to worry which side the lucky captain will choose.

A typical complaint against the proposal could be why complicate a simple toss. However, cricket has already been open to more complicated methods such as Duckworth-Lewis-Stern to make other situations fair.  If the Unlucky Captain truly believes that it does not matter who bats first, then s/he can stick to the default of zero bonus runs. Otherwise, the captain can do the following thought experiment. “Would I prefer to bowl or bat? Would I be indifferent if I added 10 runs to the lesser preferred option? Or 20 runs?” At some point, one would be indifferent between the two options because of the bonus runs added to the lesser preferred option. Cricket legend Sunil Gavaskar opined that that issue of the unfair toss ``i\emph{s something for the ICC Cricket Committee to get their heads around and make sure that there is a level playing field for both teams}''~\citep{Nand21a}. Ian Chappel echoed similar thoughts~\citep{Hind21a}. The main objective of the Toss, Propose and Choose method is to level the playing field and eliminate the potentially significant headstart the winner of the coin toss gets. 

\section{Comparison with Other Proposals}

Apart from my proposal, a few other ideas have been floated to avoid the undue impact of a coin. A frequently proposed solution is for the teams to alternate batting first over the course of a series. The solution regains some fairness over time, especially if the series has enough matches. However, it doesn’t resolve the issue of an unfair toss for the deciding or final match or for tournaments. Another proposal is for the weaker, or the touring team, to decide. However, the proposal may give too much advantage to the slightly weaker team and may incentivize teams to have lower rankings. Secondly, the match (say the ICC World Cup) may be on neutral ground, in which case both teams may be touring. 

One elegant proposal to achieve fairness is to get rid of the toss entirely and introduce an auction~\citep{Fran18a,SoDe21a}. It leads to certain issues such as ensuring simultaneous bids in a transparent yet TV-friendly manner; the decision always going to the more aggressive bidder (that helps teams that are one-trick ponies such as always chasing); the higher bidder being victim of the winner’s curse (a phenomenon that is encountered in real estate auctions); and the need for tie-breaking under identical bids. Finally, bypassing the toss completely eliminates an element that many might see as intrinsic to traditional cricketing sensibilities. In this sense, the Toss, Propose and Choose method is a minimal and transparent adaptation of the existing toss to regain fairness and only requires a handicap elicitation from one of the captains. It makes it easy to understand why both sides are free of envy of each other. If the Unlucky Captain is indifferent between bowling and batting first, the method coincides with the traditional toss. In 2018, some of the fair alternatives were considered but the ICC decided to keep the toss~\citep{Wisd18a}. The decision was in line with the MCC clause 13.5 (``''\emph{the captain of the side winning the toss shall decide whether to bat or to field and shall notify the opposing captain and the umpires of this decision}''~\citep{MCC}). My proposal is to keep the toss but to make it fair. 

\section{Final Words}

So is the cricket community ready to consider the Toss, Propose and Choose method that is designed for a high-stakes match? Ironically, I feel that a suitable place to experiment is in low-stakes junior matches to see how the cricketing community feels about it. I won’t complain if the proposal also gets taken up at the next ICC Test Championship Final! I suspect that the captain who loses the toss won’t either.

\section*{Acknowledgments}

The author thanks Robbie Boland for valuable comments. 
 


 \bibliographystyle{aer2} 
 
 \ifx\undefined\bysame
 \newcommand{\bysame}{\leavevmode\hbox to\leftmargin{\hrulefill\,\,}}
 \fi

	\end{document}